# Harmonic generation and energy transport in dielectric and semiconductors at visible and UV wavelengths: the case of GaP

**V. Roppo[1,2], N. Akozbek[3], D. de Ceglia[3], M.A. Vincenti[3], M. Scalora[2,*]**

[1] *Universitat Politècnica de Catalunya, Departament de Física i Enginyeria Nuclear, Rambla Sant Nebridi, 08222 Terrassa, Spain*

[2] *Charles M. Bowden Research Center, RDECOM, Redstone Arsenal, AL 35898, USA*

[3] *AEgis Technologies Group, 410 Jan Davis Dr., Huntsville, AL 35806, USA*

[*] corresponding author: michael.scalora@us.army.mil

**Abstract**

We study inhibition of absorption, transparency, energy and momentum transport of the inhomogeneous component of harmonic pulses in dielectrics and semiconductors, at visible and UV wavelengths, focusing on materials like GaP. In these spectral regions GaP is characterized by large absorption, metallic behavior or a combination of both. We show that phase locking causes the generated inhomogeneous signals to propagate through a bulk metallic medium without being absorbed, that is occurs even in centrosymmetric materials via the magnetic Lorentz force, and that the transport of energy and momentum is quite peculiar and seemingly anomalous. These results make it clear that there are new opportunities in ultrafast nonlinear optics and nano-plasmonics in new wavelength ranges.

## Introduction

Harmonic generation is a nonlinear process that has also been used to generate new coherent light sources towards shorter wavelengths (for a recent review see [1]). Compact coherent sources particularly in the extreme ultraviolet (EUV) spectral region have important applications ranging from nanotechnology and high resolution microscopy to spectroscopy [1]. Most nonlinear optical materials absorb strongly in the UV region ($\lambda \leq 400$nm). For this reason harmonic generation has been limited to visible wavelengths to avoid absorption losses. Absorption is generally considered detrimental because it can limit the conversion efficiency over the coherence length. In fact, one reasonably expects that any generated harmonic will be fully re-absorbed inside the medium if the sample is much longer than the characteristic absorption and coherence lengths.

In transparent materials the general solution for second and third harmonic generation (SHG and THG) from a boundary layer consists of a reflected signal and two forward-propagating components. The latter displays k-vectors that are solutions of the homogenous and inhomogeneous wave equations [2,3]. For example, Maker fringes result from the interference between the two forward propagating components that at phase matching merge into a single solution [4]. Recently, complete phase and group velocity locking were demonstrated in a $LiNbO_3$ wafer, for different combinations of incident polarizations [5].

The generation and transmission of SH and TH wavelengths was recently demonstrated both theoretically and experimentally in bulk GaAs in a case where the harmonics were tuned well above the absorption band edge [6]. The phenomenon was attributed to phase locking between the fundamental and its generated inhomogeneous harmonics that occurs under phase mismatched conditions, irrespective of material parameters [7], as long as the material is somewhat transparent to the pump. The pump then impresses its dispersive properties to its harmonics, which in turn experience no absorption or other dephasing effects such as group velocity walk-off. The harmonics co-propagate phase- and velocity-locked to the pump, as energy exchange ceases away from interfaces [7]. In cavity situations phase locking persists in a GaAs etalon [8] and in a GaAs defect layer surrounded by Bragg mirrors [9]. Even though the generated signals were tuned far below the absorption edge of GaAs (612nm and 408nm), conversion efficiencies were shown to improve by nearly four orders of magnitudes compared to bulk GaAs. The generated inhomogeneous signals thus compete with the usual homogeneous signals that are much more abundant at the phase matching condition.

Perhaps more relevant for our purposes are the theoretical prediction and experimental verification that induced transparency occurs also in generally forbidden ranges, i.e. where the dielectric constant of materials displays metallic behavior (like GaP at 223nm) [10]. The objective of this work is to perform a more detailed theoretical study of energy and momentum transport in GaP such that the harmonics are tuned in ranges that are either strongly absorptive or where the dielectric constant becomes negative, yielding metallic behavior. In both cases the homogeneous SH and TH signals vanish, leaving the inhomogeneous components unscathed. This concept is crucial to appreciate the dynamics that underlie the interaction. In reference [11] it was shown that the one dimensional field localization patterns change sensibly if the absorption of the material is turned on or off, because the generated signal is generally a superposition of homogeneous and inhomogeneous signals. In a two-dimensional environment the absence of absorption can cause the momentum vector of a given harmonic to oscillate in direction and amplitude as energy flows back and forth between the two generated components until walk-off separates them. This simple realization makes it possible to use momentum and energy flow as yet another degree of freedom exploitable in the design of nonlinear devices.

Absorption and momentum direction are two physical characteristics of matter-wave interaction that seldom come under scrutiny, especially in metals and semiconductors at visible and UV wavelengths. Bulk metals have their plasma frequencies in the UV-visible region due to their relatively large free carrier densities, and exhibit a negative permittivity from the visible all the way to microwave frequencies. Unlike bulk metals, however, semiconductors like GaAs, GaP, Si, and Ge exhibit a region of negative dielectric permittivity at deep UV wavelengths and are transparent in the infrared region. Negative refraction that is generally associated with negative index materials ($\varepsilon<0$ and $\mu<0$) occurs also for a TM-polarized field incident on a negative permittivity medium [12]. This has led to the development of a tunable, high-transmission superlens in the visible range using resonant metal/dielectric multilayer structures [13, 14]. The same arrangement may be exploited in resonant multilayer stacks such as GaAs/KCl and GaAs/MgO operating in the deep UV spectral region [15], and a GaAs grating to induce enhanced transmission of light also at UV wavelengths [16]. These structures may provide an alternative in the experimental realization of nonlinear optical and plasmonic phenomena in the UV-Visible range since the fabrication of bulk, effective negative index materials has proven to be challenging due to absorption losses. Absorption is still the major limiting factor for any practical applications, and various gain schemes have been considered. Semiconductors are used extensively in photonics and electronics industry and are easier to integrate with current electro-optical technologies. They can also be doped to control free carrier density and plasma frequency. In fact, doped semiconductors that exhibit a negative permittivity in the far IR region were shown to support negative refraction [17], plasmonic response [18], and superlensing in SiC [19]. In addition, semiconductors exhibit large nonlinear coefficients, an important factor for nonlinear frequency conversion and other types of parametric processes. The question we address presently may be formulated as follows: How does the transport of energy and momentum occur for the pump and the harmonics tuned in ranges of high absorption and where ε<0, μ=1?

## Theoretical model

In what follows we present the model we use to describe ultrashort pulse propagation phenomena that include SHG and THG in GaP (data from reference [20]). The study is completely generic and the same strategy can be applied to other dielectrics and/or semiconductors. We assume the medium is composed of a collection of doubly-resonant Lorentz oscillators (the double resonance system is clearly visible in the linear dielectric functions shown in Fig.1.)

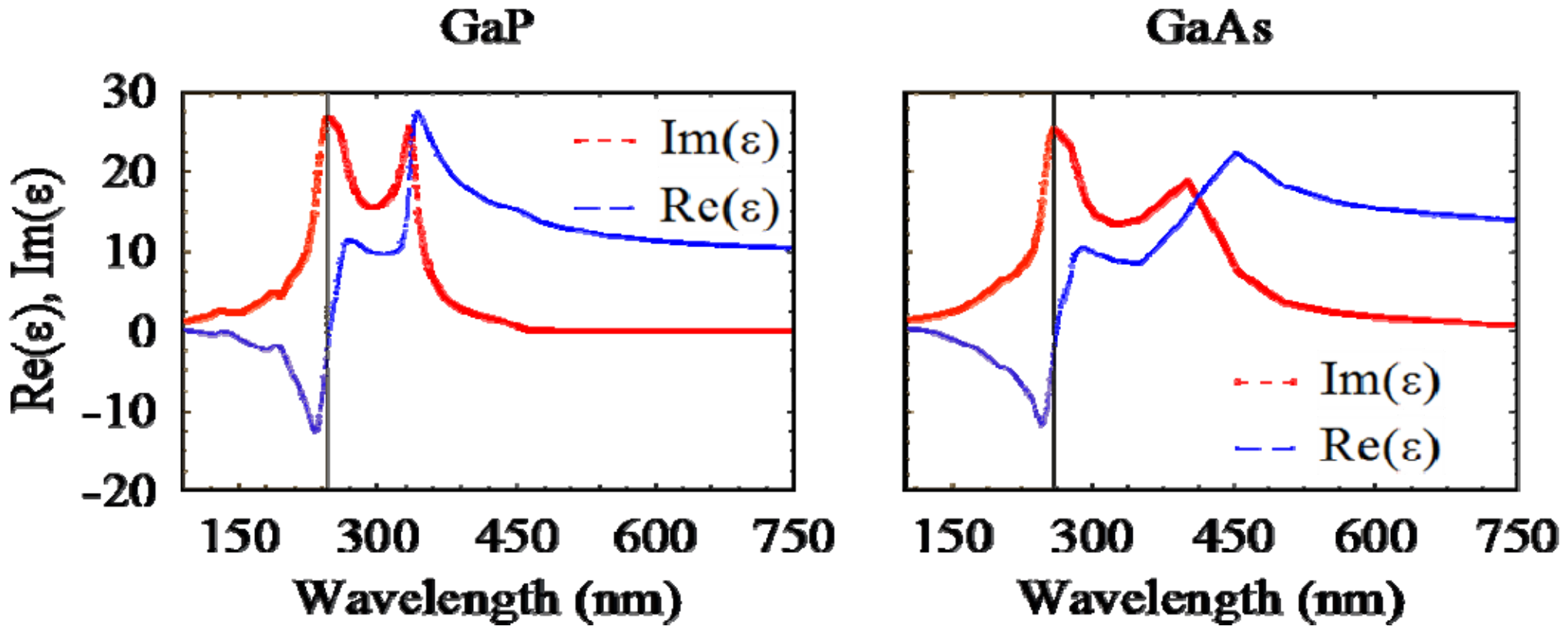

**Fig.1.** Dielectric constant of GaP (left) and GaAs (right), taken from reference [21]. The regions of negative dielectric constant (~100nm-250nm) are denoted by the shaded areas.

With reference to Fig.2, we assume that both TE- and TM-polarized fields may be present and are decomposed as a superposition of harmonics as follows:

$$\mathbf{E}=\begin{pmatrix}E_x\\E_y\\E_z\end{pmatrix}=\begin{pmatrix}\mathbf{i}\left(E^{\omega}_{TEx}e^{-i\omega t}+\left(E^{\omega}_{TEx}\right)^*e^{i\omega t}+E^{2\omega}_{TEx}e^{-2i\omega t}+\left(E^{2\omega}_{TEx}\right)^*e^{2i\omega t}+E^{3\omega}_{TEx}e^{-3i\omega t}+\left(E^{3\omega}_{TEx}\right)^*e^{3i\omega t}\right)\\+\mathbf{j}\left(E^{\omega}_{TMy}e^{-i\omega t}+\left(E^{\omega}_{TMy}\right)^*e^{i\omega t}+E^{2\omega}_{TMy}e^{-2i\omega t}+\left(E^{2\omega}_{TMy}\right)^*e^{2i\omega t}+E^{3\omega}_{TMy}e^{-3i\omega t}+\left(E^{3\omega}_{TMy}\right)^*e^{3i\omega t}\right)\\+\mathbf{k}\left(E^{\omega}_{TMz}e^{-i\omega t}+\left(E^{\omega}_{TMz}\right)^*e^{i\omega t}+E^{2\omega}_{TMz}e^{-2i\omega t}+\left(E^{2\omega}_{TMz}\right)^*e^{2i\omega t}+E^{3\omega}_{TMz}e^{-3i\omega t}+\left(E^{3\omega}_{TMz}\right)^*e^{3i\omega t}\right)\end{pmatrix} \quad (1)$$

$$\mathbf{H}=\begin{pmatrix}H_x\\H_y\\H_z\end{pmatrix}=\begin{pmatrix}\mathbf{i}\left(H_{TMx}^{\omega}e^{-i\omega t}+\left(H_{TMx}^{\omega}\right)^*e^{i\omega t}+H_{TMx}^{2\omega}e^{-2i\omega t}+\left(H_{TMx}^{2\omega}\right)^*e^{2i\omega t}+H_{TMx}^{3\omega}e^{-3i\omega t}+\left(H_{TMx}^{3\omega}\right)^*e^{3i\omega t}\right)\\+\mathbf{j}\left(H_{TEy}^{\omega}e^{-i\omega t}+\left(H_{TEy}^{\omega}\right)^*e^{i\omega t}+H_{TEy}^{2\omega}e^{-2i\omega t}+\left(H_{TEy}^{2\omega}\right)^*e^{2i\omega t}+H_{TEy}^{3\omega}e^{-3i\omega t}+\left(H_{TEy}^{3\omega}\right)^*e^{3i\omega t}\right)\\+\mathbf{k}\left(H_{TEz}^{\omega}e^{-i\omega t}+\left(H_{TEz}^{\omega}\right)^*e^{i\omega t}+H_{TEz}^{2\omega}e^{-2i\omega t}+\left(H_{TEz}^{2\omega}\right)^*e^{2i\omega t}+H_{TEz}^{3\omega}e^{-3i\omega t}+\left(H_{TEz}^{3\omega}\right)^*e^{3i\omega t}\right)\end{pmatrix}. \quad (2)$$

Then, a TM-polarized field has a magnetic field component along x and electric fields that point along y and z. A TE-polarized field has magnetic field components along the y- and z-directions, and a single electric field component that points in the x-direction. The second order polarization vector of GaP (or GaAs) may be written as follows [22]:

$$\begin{pmatrix}P_{NL,x}^{(2)}\\P_{NL,y}^{(2)}\\P_{NL,z}^{(2)}\end{pmatrix}=2\mathrm{d}_{14}\begin{pmatrix}E_yE_z\\E_x\,E_z\\E_xE_y\end{pmatrix}, \quad (3)$$

where the coordinates correspond to the crystals' principal axes. Substituting the E-field vector defined in Eq.2 into Eq.4 leads to the following equations:

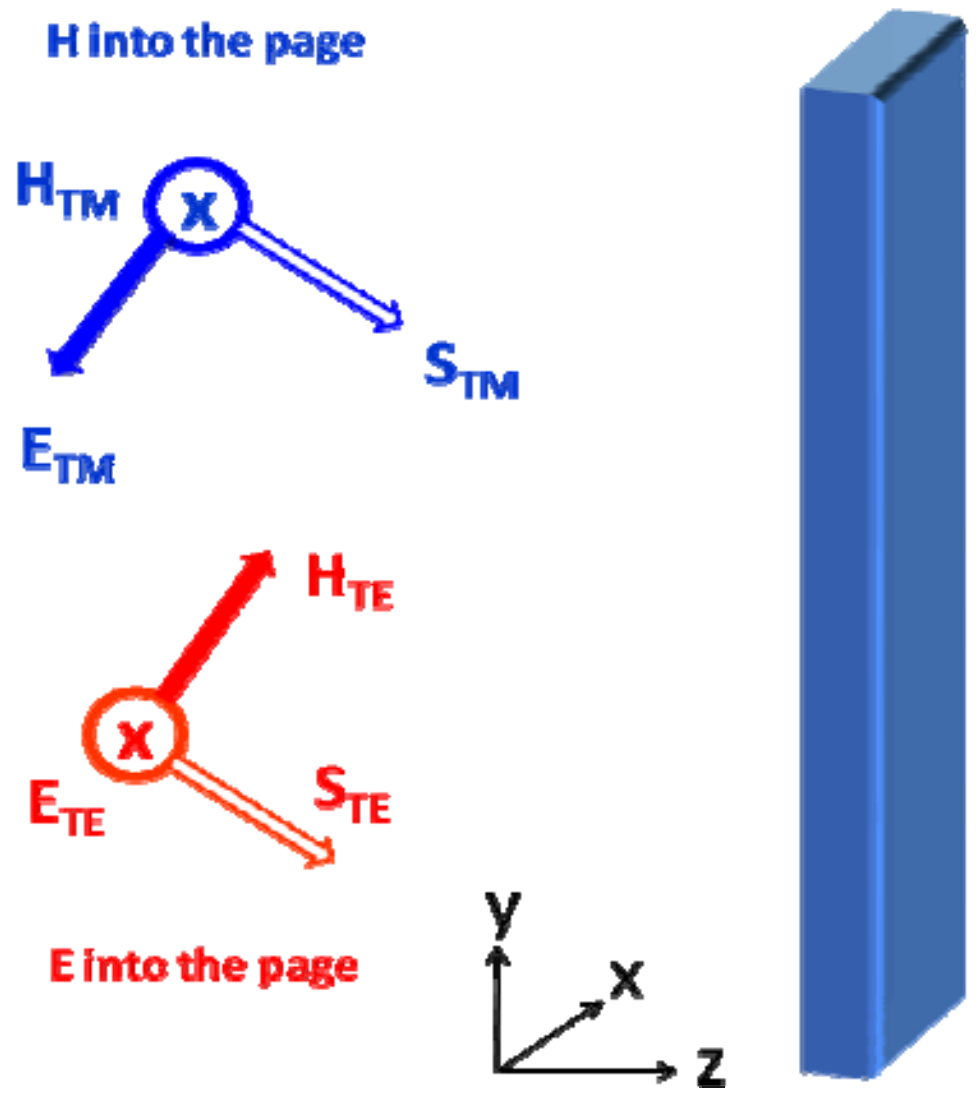

**Fig.2.** Generic scheme of the interaction. The incident fields need not necessarily be collinear.

$$P_{NL,x}^{(2)} = 2\mathrm{d}_{14}\begin{pmatrix} \left( E_{TMy}^{2\omega}\left(E_{TMz}^{\omega}\right)^{*} + \left(E_{TMy}^{\omega}\right)^{*} E_{TMz}^{2\omega} + E_{TMy}^{3\omega}\left(E_{TMz}^{2\omega}\right)^{*} + \left(E_{TMy}^{2\omega}\right)^{*} E_{TMz}^{3\omega} \right) e^{-i\omega t} \\ + \left( E_{TMy}^{\omega} E_{TMz}^{\omega} + E_{TMy}^{3\omega}\left(E_{TMz}^{\omega}\right)^{*} + \left(E_{TMy}^{\omega}\right)^{*} E_{TMz}^{3\omega} \right) e^{-2i\omega t} \\ + \left( E_{TMy}^{2\omega} E_{TMz}^{\omega} + E_{TMy}^{\omega} E_{TMz}^{2\omega} \right) e^{-3i\omega t} \end{pmatrix} \tag{4}$$

$$P_{NL,y}^{(2)} = 2\mathrm{d}_{14}\begin{pmatrix} \left( E_{TEx}^{2\omega}\left(E_{TMz}^{\omega}\right)^{*} + \left(E_{TEx}^{\omega}\right)^{*} E_{TMz}^{2\omega} + E_{TEx}^{3\omega}\left(E_{TMz}^{2\omega}\right)^{*} + \left(E_{TEx}^{2\omega}\right)^{*} E_{TMz}^{3\omega} \right) e^{-i\omega t} \\ + \left( E_{TEx}^{\omega} E_{TMz}^{\omega} + E_{TEx}^{3\omega}\left(E_{TMz}^{\omega}\right)^{*} + \left(E_{TEx}^{\omega}\right)^{*} E_{TMz}^{3\omega} \right) e^{-2i\omega t} \\ + \left( E_{TEx}^{2\omega} E_{TMz}^{\omega} + E_{TEx}^{\omega} E_{TMz}^{2\omega} \right) e^{-3i\omega t} \end{pmatrix} \tag{5}$$

$$P_{NL,z}^{(2)} = 2\mathrm{d}_{14}\begin{pmatrix} \left( E_{TEx}^{2\omega}\left(E_{TMy}^{\omega}\right)^{*} + \left(E_{TEx}^{\omega}\right)^{*} E_{TMy}^{2\omega} + E_{TEx}^{3\omega}\left(E_{TMy}^{2\omega}\right)^{*} + \left(E_{TEx}^{2\omega}\right)^{*} E_{TMy}^{3\omega} \right) e^{-i\omega t} \\ + \left( E_{TEx}^{\omega} E_{TMy}^{\omega} + E_{TEx}^{3\omega}\left(E_{TMy}^{\omega}\right)^{*} + \left(E_{TEx}^{\omega}\right)^{*} E_{TMy}^{3\omega} \right) e^{-2i\omega t} \\ + \left( E_{TEx}^{2\omega} E_{TMy}^{\omega} + E_{TEx}^{\omega} E_{TMy}^{2\omega} \right) e^{-3i\omega t} \end{pmatrix} \tag{6}$$

From Eqs.4-6 alone one may surmise that if a TM-polarized field were incident on GaP it would generate a TE-polarized SH signal via the term $E_{TEx}^{2\omega} \sim \mathbf{i}\left(E_{TMy}^{\omega} E_{TMz}^{\omega}\right)e^{-2i\omega t}$ in Eq.5. In turn, together with the TM-polarized pump, a non-zero $E_{TEx}^{2\omega}$ triggers TM-polarized, cascaded THG via the terms: $E_{TMy}^{3\omega} \sim \mathbf{j}\left(E_{TEx}^{2\omega} E_{TMz}^{\omega}\right)e^{-3i\omega t}$ and $E_{TMz}^{3\omega} \sim \mathbf{k}\left(E_{TEx}^{2\omega} E_{TMy}^{\omega}\right)e^{-3i\omega t}$ in Eqs.(5) and (6), respectively. Finally, some photons are returned to the original TM-polarized pump via a down-conversion process by the terms: $E_{TMy}^{\omega} \sim \mathbf{j}\left(E_{TEx}^{2\omega}\left(E_{TMz}^{\omega}\right)^{*} + \left(E_{TEx}^{2\omega}\right)^{*} E_{TMz}^{3\omega}\right)e^{-i\omega t}$ and $E_{TMz}^{\omega} \sim \mathbf{k}\left(E_{TEx}^{2\omega}\left(E_{TMy}^{\omega}\right)^{*} + \left(E_{TEx}^{2\omega}\right)^{*} E_{TMy}^{3\omega}\right)e^{-i\omega t}$.

Since a TM-polarized SH component is notably absent from the predictions made using the $\chi^{(2)}$ of Eq.(3), in the absence of other sources this portion of the SH signal must originate in surface and volume nonlinear phenomena ordinarily associated with centrosymmetric materials [10,21]. Like all materials, GaP is also characterized by surface and volume nonlinear sources arising from symmetry breaking and from the magnetic portion of the Lorentz force, independently of its bulk $\chi^{(2)}$ and $\chi^{(3)}$ properties. These contributions are taken into account by deriving equations of motion for the components of the polarization of bound electrons beginning from a classical Lorentz oscillator model [15,16,21]:

$$\ddot{\mathbf{P}}_{b,\omega} + \tilde{\gamma}_{b,\omega}\dot{\mathbf{P}}_{b,\omega} + \tilde{\omega}_{0,b,\omega}^2\mathbf{P}_{b,\omega} \approx \frac{n_{0,b}e^2\lambda_0^2}{m_b^*c^2}\mathbf{E}_\omega + \frac{e\,\lambda_0}{m_b^*c^2}\begin{pmatrix} -\frac{1}{2}\mathbf{E}_\omega^*\nabla\bullet\mathbf{P}_{b,2\omega} \\ +2\mathbf{E}_{2\omega}\nabla\bullet\mathbf{P}_{b,\omega}^* \\ -\frac{2}{3}\mathbf{E}_{2\omega}^*\nabla\bullet\mathbf{P}_{b,3\omega} \\ -\frac{3}{2}\mathbf{E}_{3\omega}\nabla\bullet\mathbf{P}_{b,2\omega}^* \end{pmatrix} + \frac{e\,\lambda_0}{m_b^*c^2}\begin{pmatrix} \left(\dot{\mathbf{P}}_{b,\omega}^* + i\omega\mathbf{P}_{b,\omega}^*\right)\times\mathbf{H}_{2\omega} \\ +\left(\dot{\mathbf{P}}_{b,2\omega} - 2i\omega\mathbf{P}_{b,2\omega}\right)\times\mathbf{H}_\omega^* \\ +\left(\dot{\mathbf{P}}_{b,2\omega}^* + 2i\omega\mathbf{P}_{b,2\omega}^*\right)\times\mathbf{H}_{3\omega} \\ +\left(\dot{\mathbf{P}}_{b,3\omega} - 3i\omega\mathbf{P}_{b,3\omega}\right)\times\mathbf{H}_{2\omega}^* \end{pmatrix}, \tag{7}$$

$$\ddot{\mathbf{P}}_{b,2\omega} + \tilde{\gamma}_{b,2\omega}\dot{\mathbf{P}}_{b,2\omega} + \tilde{\omega}_{0,b,2\omega}^2\mathbf{P}_{b,2\omega} \approx \frac{n_{0,b}e^2\lambda_0^2}{m_b^*c^2}\mathbf{E}_{2\omega} + \frac{e\,\lambda_0}{m_b^*c^2}\begin{pmatrix} \mathbf{E}_\omega\nabla\bullet\mathbf{P}_{b,\omega} \\ -\frac{1}{3}\mathbf{E}_\omega^*\nabla\bullet\mathbf{P}_{b,3\omega} \\ -3\mathbf{E}_{3\omega}\nabla\bullet\mathbf{P}_{b,\omega}^* \end{pmatrix} + \frac{e\,\lambda_0}{m_b^*c^2}\begin{pmatrix} \left(\dot{\mathbf{P}}_{b,\omega} - i\omega\mathbf{P}_{b,\omega}\right)\times\mathbf{H}_\omega \\ +\left(\dot{\mathbf{P}}_{b,\omega}^* + i\omega\mathbf{P}_{b,\omega}^*\right)\times\mathbf{H}_{3\omega} \\ +\left(\dot{\mathbf{P}}_{b,3\omega} - 3i\omega\mathbf{P}_{b,3\omega}\right)\times\mathbf{H}_\omega^* \end{pmatrix}, \tag{8}$$

$$\ddot{\mathbf{P}}_{b,3\omega} + \tilde{\gamma}_{b,3\omega}\dot{\mathbf{P}}_{b,3\omega} + \tilde{\omega}_{0,b,3\omega}^2\mathbf{P}_{b,3\omega} \approx \frac{n_{0,b}e^2\lambda_0^2}{m_b^*c^2}\mathbf{E}_{3\omega} + \frac{e\,\lambda_0}{m_b^*c^2}\begin{pmatrix} \frac{1}{2}\mathbf{E}_\omega\nabla\bullet\mathbf{P}_{b,2\omega} \\ +2\mathbf{E}_{2\omega}\nabla\bullet\mathbf{P}_{b,\omega} \end{pmatrix} + \frac{e\,\lambda_0}{m_b^*c^2}\begin{pmatrix} \left(\dot{\mathbf{P}}_{b,2\omega} - 2i\omega\mathbf{P}_{b,2\omega}\right)\times\mathbf{H}_\omega \\ +\left(\dot{\mathbf{P}}_{b,\omega} - i\omega\mathbf{P}_{b,\omega}\right)\times\mathbf{H}_{2\omega} \end{pmatrix}. \tag{9}$$

In deriving Eqs.(7-9) we have neglected higher order quadrupole-like terms. The scaled coefficients are $\tilde{\gamma}_{b,N\omega} = (\gamma_b - Ni\omega)$, $\tilde{\omega}_{0,N\omega}^2 = (\omega_{0,b}^2 - (N\omega)^2 + i\gamma_b N\omega)$, where N is an integer that denotes the given harmonic order. $\lambda_0 = 1\mu\text{m}$ is the reference wavelength; $\mathbf{P}_{b,N\omega}$ is the polarization envelope of the Nth harmonic; *c* is the speed of light in vacuum; the subscript *b* stands for bound; *e* is the electron charge; $m_b^*$ is the effective mass of bound electrons. Nonlinear source terms in Eqs.7-9 (bound charges in GaP via quadrupole-like contributions proportional to $\nabla\bullet\mathbf{P}_{b,N\omega}$, and magnetic terms like $\left(\dot{\mathbf{P}}_{b,\omega} - i\omega\mathbf{P}_{b,\omega}\right)\times\mathbf{H}_\omega$) give rise to TM-polarized SH and TH fields. With this in mind another look at Eqs.4-6 reveals that the TM-polarized pump and its harmonics serve as nonlinear sources for all TE-polarized fields, including the pump [15,16,21]. The production of a TE-polarized pump field initiates with the introduction of either symmetry breaking (i.e. $\nabla \bullet \mathbf{P}_{b,N\omega}$, surface terms) or the Lorentz force. Once TE-polarized fields are generated, all interaction channels become active and the generation of all harmonic fields takes place. We should note that this is a peculiarity of the $\chi^{(2)}$ tensor of GaP and other materials that have the same symmetry properties [22]. For further details about the model and the integration scheme employed we direct the reader to references [15,16,21].

The inclusion of third order phenomena begins with the general expansion of the third order polarization as follows [22]:

$$P_{NL,i}^{(3)} = \sum_{j=1,3} \sum_{k=1,3} \sum_{l=1,3} \chi_{i,j,k,l}^{(3)} E_j E_k E_l \qquad j,k,l = x,y,z \tag{10}$$

GaP has cubic symmetry of the type $\overline{4}3m$, so that Eq.6 reduces to:

$$\begin{aligned} P_{NL,x}^{(3)} &= \chi_{xxxx}^{(3)} E_x^3 + 3\chi_{xxyy}^{(3)} E_y^2 E_x + 3\chi_{xxzz}^{(3)} E_z^2 E_x \\ P_{NL,y}^{(3)} &= \chi_{yyyy}^{(3)} E_y^3 + 3\chi_{xxyy}^{(3)} E_x^2 E_y + 3\chi_{yyzz}^{(3)} E_z^2 E_y \\ P_{NL,z}^{(3)} &= \chi_{zzzz}^{(3)} E_z^3 + 3\chi_{zzxx}^{(3)} E_x^2 E_z + 3\chi_{zzyy}^{(3)} E_y^2 E_z \end{aligned} \tag{11}$$

Substituting Eqs.1-2 into Eqs.11 leads to $\chi^{(3)}$ contributions to all harmonic components, with self- and cross-phase modulation along with terms that couple orthogonal polarization states. In particular, the presence of non-zero second harmonic fields provides a way for the $\chi^{(3)}$ to also contribute to the SHG process. Finally, the linear and nonlinear polarization components at each frequency are gathered and combined into Maxwell's equations.

## Numerical results

We have already seen in the previous section that important results can be deduced by analyzing the equations of motion to deduce the possible energy fluxes that can be activated among the fundamental and the generated harmonics. Further details about the dynamics may be ascertained by numerically solving Maxwell's system of equations coupled to the nonlinear polarizations exemplified by Eqs.(4-11). In our examples we use a pulse approximately 30fs in duration tuned to 670nm propagating across an air-GaP interface [10]. The SH and TH are thus tuned at 335nm and 223nm respectively. The dispersion values of the permittivity at the fundamental, SH and TH are taken as $\varepsilon_{FF}$=11.12, $\varepsilon_{SH}$=11.96 + i24 and $\varepsilon_{TH}$=-12.8 + i 9.73, respectively

In Fig.3 we report a snapshot of the dynamics as the incident TM-polarized pulse generates TE- and TM-polarized harmonics that in turn generate down-converted TE-polarized pump photons in the manner previously described. The media file in Fig.3 shows that all the generated pulses are locked to the pump during the entire process and propagate in the same direction, including the TM-polarized SH field. By observing the color scale of each harmonic and comparing it to the TM-polarized pump scale one can ascertain the approximate conversion efficiency of a given harmonic. The figure suggests that the generation of TE-polarized pump photons is similar to TM-polarized THG and several orders of magnitude larger than TE-polarized THG. The process is nevertheless more difficult to observe because one needs a large degree of discrimination between the intense TM-polarized pump and the much weaker TE-polarized signal. What makes this process unusual, unique, and perhaps exploitable for new types of applications is that *phase locking occurs even in the absence of a nonlinear bulk coefficient, via the magnetic Lorentz Force, as evidenced by the action of the TM-polarized SH signal*. This is confirmed by performing the simulation with second and third order nonlinear coefficients set to zero.

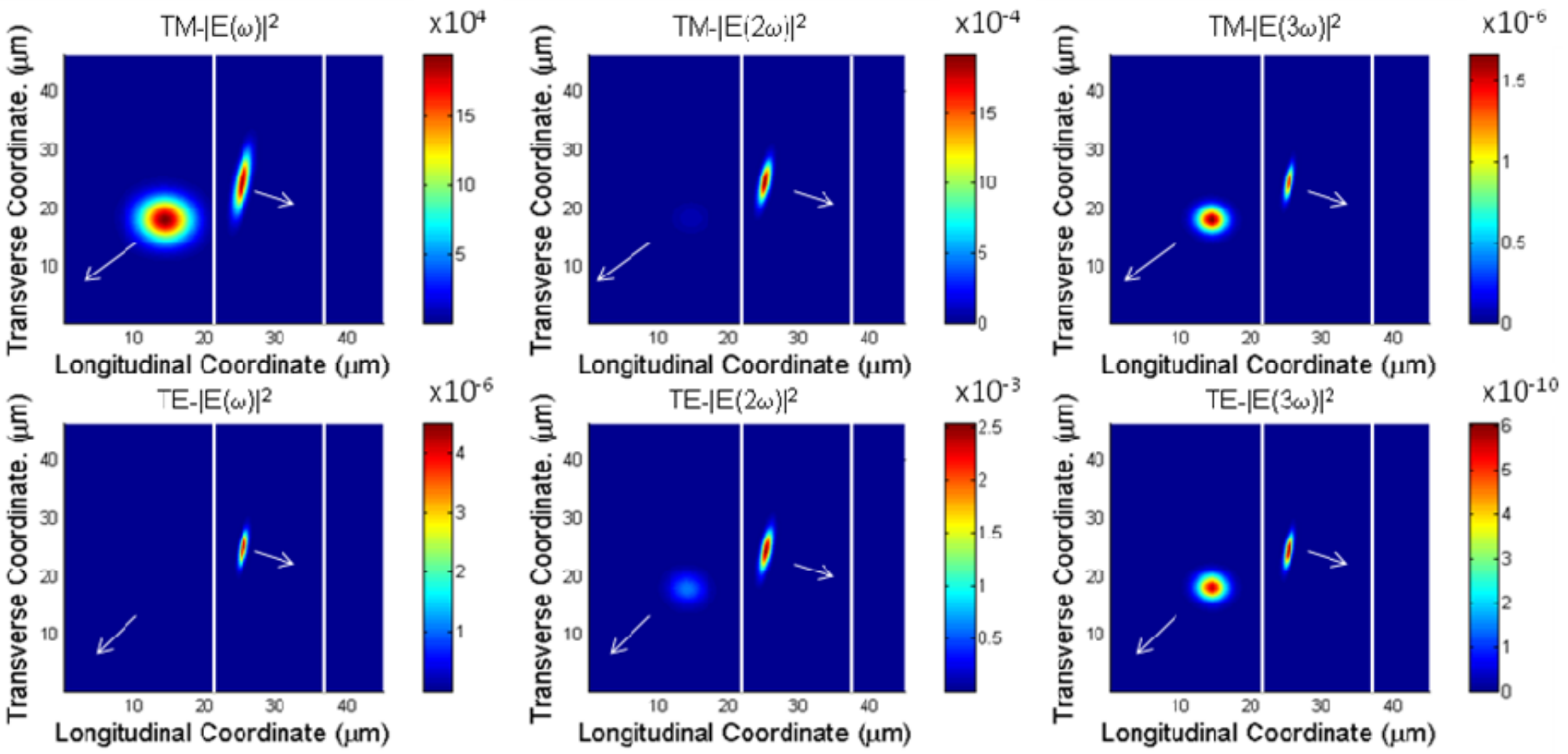

**Fig.3 (Media 1)** A TM-polarized pump pulse ~30fs in duration tuned to 670nm crosses into GaP (denoted by white boundaries) and generates SH and TH fields. (Top row) TM polarized fields. (Bottom row) TE polarized fields. Pump refraction into the medium proceeds as predicted by Snell's law. Part of the signals is reflected specularly and partly is transmitted, with similar results for all generated fields.

In that case, only TM-polarized fields are generated in a manner nearly identical to that of Fig.3. Suffice it to say for the moment that SHG in centrosymmetric dielectric materials like Silicon, for example, may yield unexpected benefits even in cavity environments because phase locking persists with the mere presence of the Lorentz magnetic force, as our calculations show.

In Fig.4 we summarize the results by highlighting the directions of the individual light momenta for the pump and each of the generated fields. Fig.4a corresponds to the situation of Fig.3. In Fig.4b the simulation is repeated for an incident field tuned to 1200nm, yielding comparable results. A comparison of the propagation event in the media file in Fig.3 and the scheme in Fig.4a shows that there is an apparent disparity between the directions of propagation of the TM-polarized SH pulse and its associated momentum vector. For example, in Fig.3 the TM-polarized harmonic is seen to clearly follow the path of the pump pulse, while its momentum points in a completely different direction (fig.4a). To reconcile these apparent differences we calculate the electromagnetic momentum of a wave packet located inside a medium of thickness $L$ as a function of time using the usual expression:

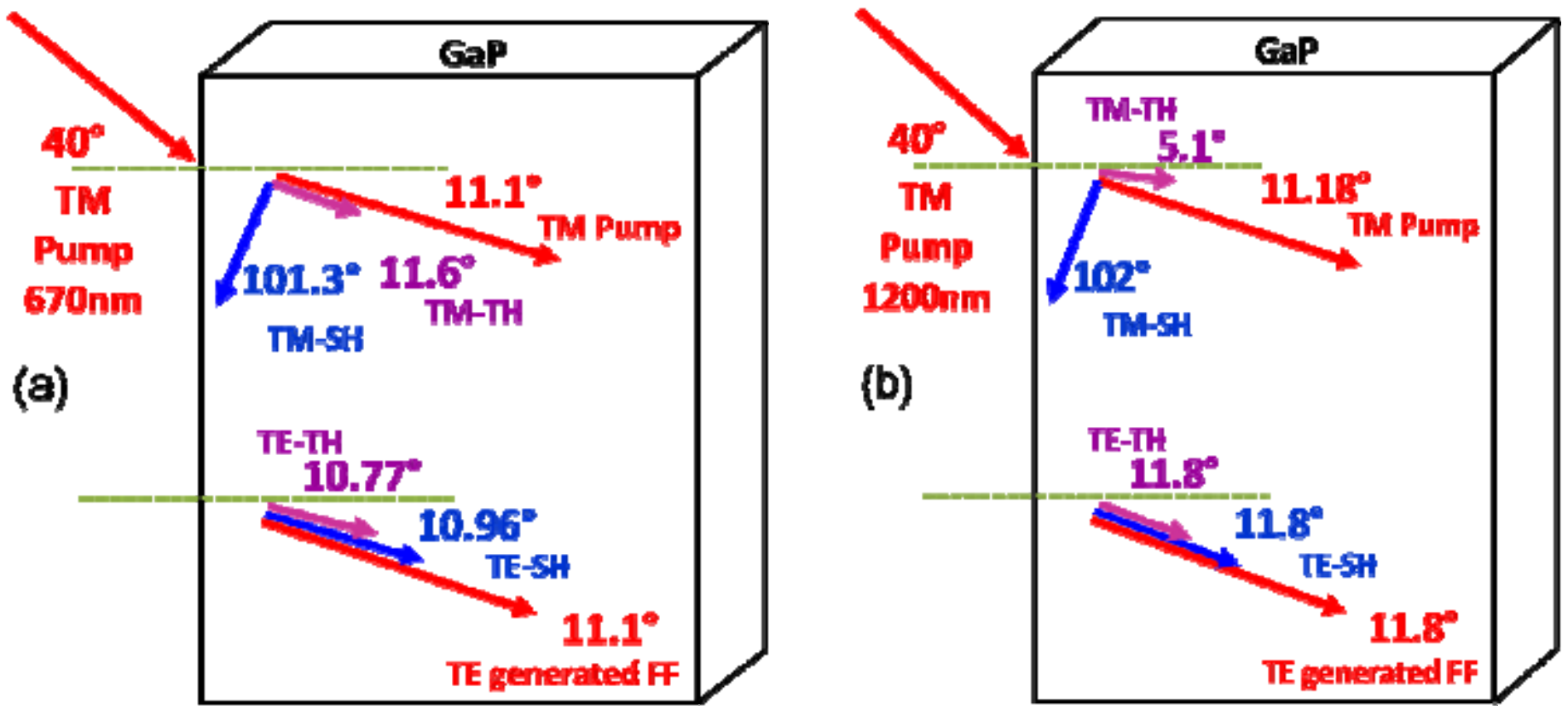

**Fig.4**. Pump (red), SH (blue) and TH (magenta) momentum refraction angles. The pump's momentum refraction angles coincide well with predictions made using Snell's law. Pump tuning at 670nm (a) and 1200nm (b).

$$\mathbf{P}_{\xi,\tilde{y}}(\tau) = \frac{1}{c^2}\int_{\xi=0}^{\xi=L}\int_{\tilde{y}=-\infty}^{\tilde{y}=\infty} \mathbf{S}_{\xi}(\tilde{y},\xi,\tau)\,d\tilde{y}\,d\xi\,, \qquad (12)$$

where

$$\mathbf{S}_{\xi,\tilde{y}}(\tilde{y},\xi,\tau) = \frac{c}{4\pi}\mathbf{E}\times\mathbf{H} \qquad (13)$$

is the Abraham momentum density. After substituting Eqs.1-2 into Eq.(12) one may then use the resulting components to extract the momentum refraction angles as a function of time for each frequency as the pulse enters and settles inside the medium, for example:

$$\theta_{r,\omega}(\tau) = \tan^{-1}\left[P_{\tilde{y},\omega}(\tau)/P_{\xi,\omega}(\tau)\right]. \qquad (14)$$

This approach typically yields results nearly identical to those predicted by Snell's law, even for pulses only a few optical cycles in duration [23, 24].

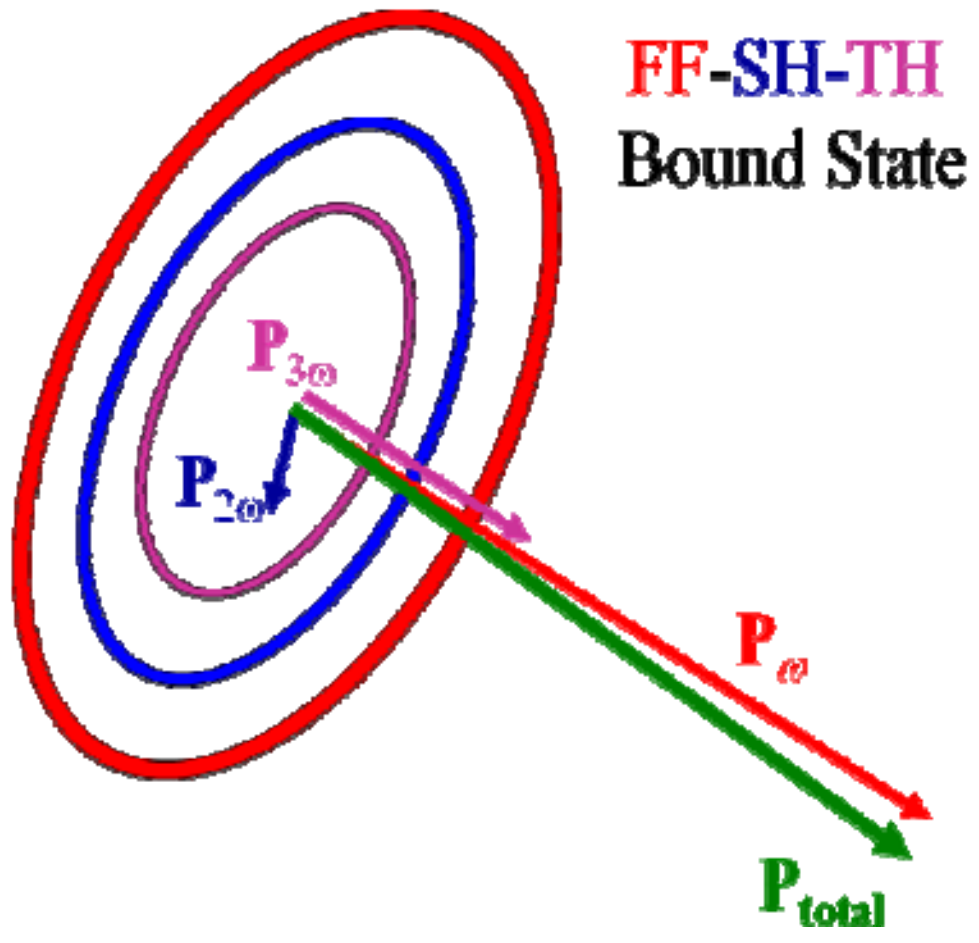

**Fig.5**. Phase locked fundamental (red shell), SH (blue shell) and TH (magenta shell) TM-polarized pulses. Pump momentum and energy nearly always overwhelm corresponding SH and TH values, so that the total momentum of the system (in green) is always approximately equal to the pump's momentum.

The apparent inconsistency between the visual clues found in the media file of Fig.3 and the results of Eq.(12) can be easily resolved because Eq.(12) in fact gives information about the total momentum of the bound pump-harmonics system, while Eq.(14) represents an arbitrary separation of the individual momenta. The situation is summarized in Fig.5. Under the conditions we have explored, a phase mismatch ensures that the pump remains undepleted for all reasonable nonlinear coefficients and incident peak powers, so that each of the momenta of the harmonic signals is always several orders of magnitude smaller compared to the pump momentum. In this view, the harmonics remain trapped by the pump and move in one direction while their momenta may actually point in another [23]. Then, the total momentum and total energy for the bound state always nearly coincide with the pump's momentum and energy. In other words, under ordinary conditions the generated harmonic pulses do not gain enough momentum/energy to significantly affect the direction of motion of the system, at least in the regime under consideration. However, one can easily create an artifact capable of depleting the pump and thus cause the momentum-energy refraction angles to depart from the prediction of Snell's law due to the intervening nonlinear interaction, making possible the observation of anomalous refraction of the generated harmonic beams.

## Conclusions

In summary, one of the unique properties of bulk semiconductors is that they exhibit a spectral range where the dielectric permittivity is negative, while providing a transparency region at near IR and longer pump wavelengths. For example, GaP is a potential candidate transparent above 500nm. While the second harmonic pulse might be tuned around the 330nm resonance of GaP, the third harmonic pulse is tuned in a region of negative permittivity [10]. These features cause the fields to behave in peculiar fashion, as phase locking forces energies flow and momenta to display apparent anomalies. In this work we have studied this dynamic theoretically and numerically. Another feature of semiconductors is that they can be doped to tune the dielectric properties [25]. It has been shown recently that nano-rings could be used to realize metamaterials in the optical regime [26]. Another approach that has been proposed involves the use of a four-level atomic medium having electric and magnetic transitions in hydrogen and neon atoms [27]. Since semiconductors already exhibit an intrinsic negative permittivity, the introduction of suitable dopant atoms could make it possible to use magnetic transitions to create an effective negative magnetic permeability. The present effort represents another tassel in a more comprehensive effort that serves to bridge the gap between a number of physical effects, such as harmonic generation, sub-diffraction limited imaging, negative refraction, and surface plasmons using nonlinear semiconductor materials.